\documentclass[10pt,a4paper,conference]{IEEEtran}
\IEEEoverridecommandlockouts
\usepackage{cite}
\usepackage{amsmath,amssymb,amsfonts}
\usepackage{algorithmic}
\usepackage{graphicx}
\usepackage{textcomp}
\usepackage{xcolor}
\usepackage{multirow}
\usepackage{hhline}
\usepackage{colortbl}
\usepackage{multicol}
\usepackage{booktabs}
\usepackage{color}   
\usepackage{newfloat}
\usepackage{mathtools}
\usepackage{float} 
\usepackage{lipsum}   
\usepackage{etoolbox}
\usepackage{array}
\usepackage{wrapfig}

\def\BibTeX{{\rm B\kern-.05em{\sc i\kern-.025em b}\kern-.08em
    T\kern-.1667em\lower.7ex\hbox{E}\kern-.125emX}}
    
\IEEEaftertitletext{\vspace{-1\baselineskip}}
\newcommand{\REM}[1]{}

\makeatletter
   \newcommand\figcaption{\def\@captype{figure}\caption}
   \newcommand\tabcaption{\def\@captype{table}\caption}

\begin{document}

\title{Mapping Stencils on Coarse-grained Reconfigurable Spatial Architecture}
\author{
\IEEEauthorblockN{
Jesmin Jahan Tithi\IEEEauthorrefmark{1}, 
Fabrizio Petrini\IEEEauthorrefmark{2}, 
Hongbo Rong\IEEEauthorrefmark{3},
Andrei Valentin\IEEEauthorrefmark{4},
Carl Ebeling\IEEEauthorrefmark{5}
}
\IEEEauthorblockA{\IEEEauthorrefmark{1}Parallel Computing Labs, Intel, Santa Clara, CA, USA. \emph{jesmin.jahan.tithi@intel.com}}
\IEEEauthorblockA{\IEEEauthorrefmark{2}Parallel Computing Labs, Intel, Santa Clara, CA, USA. \emph{fabrizio.petrini@intel.com}}
\IEEEauthorblockA{\IEEEauthorrefmark{3}Parallel Computing Labs, Intel, Santa Clara, CA, USA. \emph{hongbo.Rong@intel.com}}
\IEEEauthorblockA{\IEEEauthorrefmark{4}DCG, Intel, Santa Clara, CA, USA. \emph{valentin.m.andrei@gmail.com}}
\IEEEauthorblockA{\IEEEauthorrefmark{5}PSG, Intel, Santa Clara, CA, USA. \emph{carlebelingathome@gmail.com}}
}
\maketitle

\begin{abstract}
Stencils represent a class of computational patterns where an output grid point depends on a fixed shape of neighboring points in an input grid. Stencil computations are prevalent in scientific applications engaging a significant portion of supercomputing resources. Therefore, it has been always important to optimize stencil programs for the best performance. A rich body of research has focused on optimizing stencil computations on almost all parallel architectures.

Stencil applications have regular dependency patterns, inherent pipeline-parallelism, and plenty of data reuse. This makes these applications a perfect match for a coarse-grained reconfigurable spatial architecture (CGRA). A CGRA consists of many simple, small processing elements (PEs) connected with an on-chip network. Each PE can be configured to execute part of a stencil computation and all PEs run in parallel; the network can also be configured so that data loaded can be passed from a PE to a neighbor PE directly and thus reused by many PEs without register spilling and memory traffic. 

How to efficiently map a stencil computation to a CGRA is the key for performance.
In this paper, we show a few unique and generalizable ways of mapping one- and multi-dimensional stencil computations to a CGRA, fully exploiting the data reuse opportunities and parallelism. Our simulation experiments demonstrate that these mappings are efficient and enable the CGRA to outperform state-of-the-art GPUs.

\end{abstract}


\section{Introduction}
There is a continuing demand in many application domains for increased levels of performance and energy efficiency. Since the ``power wall" has dramatically slowed single-core processor's performance scaling, several accelerator architectures have emerged to improve performance and energy efficiency over multi-core and many-core processors in specific application domains. These architectures are often tailored using the properties inherently found in the target domain and can range in programmability (e.g., fix-function vs. fully programable vs. a partially programmable).

A coarse-grained reconfigurable spatial architecture (CGRA) consists of many small, simple yet efficient processing elements (PEs) that are laid down on a spatial grid and connected via an on-chip network. The PEs can be configured to perform a handful of operations. The network can also be configured so that data produced by a PE can be sent directly to another PE and consumed there, instead of being stored to and loaded from a shared memory. This can save significant amount of memory operations for applications with near-neighbor communications, for example, stencils, edit distance \cite{tithi2014exploiting}, and other wavefront algorithms \cite{chowdhury2017provably}.

Stencils represent a class of computational patterns prevalent in scientific applications such as fluid dynamics \cite{Spectral:Karniadakis} simulations, electromagnetic simulations \cite{Numerical}, implicit and explicit partial differential equation (PDE) iterative solvers \cite{Applied,Gauss-Seidel}, heat diffusion, and environment-simulations. Stencils typically involve an input and output grid where an output grid point is computed with some input grid points under a fixed dependency pattern. Due to the pattern, two neighboring output grid points share most of their input grid points, leading to plenty of data reuse opportunity. Since a major fraction of compute cycles of supercomputers is spent in running different stencil codes, it is important to optimize stencils for the best performance. 
 
Stencils' regular dependency patterns, inherent pipeline parallelism, and rich data-reuse opportunity make them a perfect match for a CGRA. By mapping a stencil computation to a CGRA, data can be loaded into the local storage of PEs, kept as long as needed, and directly communicated to neighbour PEs, saving a lot of register spills or trips to the shared cache and memory. This can provide significant performance boost in practice.

\underline{\bf Contributions.} We make the following contributions:
\begin{itemize}
\item We present a novel and general method of efficiently mapping one- and multi-dimensional stencils onto a CGRA. We present the data-flow graphs for several 1D and 2D stencils following our mapping technique. We also outline how a hybrid divide-and-conquer technique can be used to exploit temporal locality along with data-reuse on a CGRA.
\item We implement these mappings and evaluate them by conducting a detailed roofline analysis of performance and simulation experiments on a triggered instruction-based spatial architecture \cite{TIA:micros}. Our mappings to the CGRA outperform an optimized GPU implementation on Nvidia V100 GPU by $3\times$.
\end{itemize}
\section{Background}
In this section, we briefly introduce the concepts of CGRA and stencil.
\subsection{CGRA}
A CGRA is a computational fabric of hundreds or thousands of small processing elements (PEs) laid down on a spatial grid and connected via an on-chip network. The PEs can be programmed/configured to perform a handful of operations efficiently. The communication networks between two PEs can also be configured to create a direct producer-consumer relationship. PEs would not always need to go to shared cache or main memory to get the latest data, rather can get that from other PEs directly via the on-chip communication network. 

The CGRA optimizes byte- and word-level primitive operations, and typically includes an instruction set architecture (ISA) as its programming abstraction. Thus, PEs can be programmed by writing a sequence of software instructions. Often time, a high-level Domain Specific Language (DSL) can be defined and used to program these CGRAs as we used in this paper. Programming, simulating, and testing for CGRA is usually easier than that of FPGA which requires bit-level programming.

An algorithm for a CGRA is represented as a dataflow graph where nodes represent instructions and edges represent producer-consumer relationships between the instructions. Usually, the graph is broken into regions of pipeline stages. Each stage gets implemented by mapping the instructions to PEs, and then by connecting the PEs using the on-chip network to establish the producer-consumer relationship. For high performance, latency of each stage has to be minimized, and all available PEs should be utilized. 

The target CGRA in this paper is based on the triggered instruction architecture (TIA) \cite{TIA:micros}. The PEs and their interconnects are reconfigurable. PEs connect to each other, a scratchpad memory, and a cache hierarchy using an on-chip network. PEs have input and output queues to hold data. An instruction is mapped to a PE and is triggered as soon as all the required data in the PE's input queues are available, and the consumer PE is ready to accept the new data to be produced. The CGRA also shares a coherent shared virtual memory address space with an associated CPU process that runs the driver/host portion of the code and offloads kernels to the CGRA. Further detail can be found in \cite{TIA:micros}. 

\begin{figure}[pht!]
\centering
\includegraphics[width=0.4\textwidth]{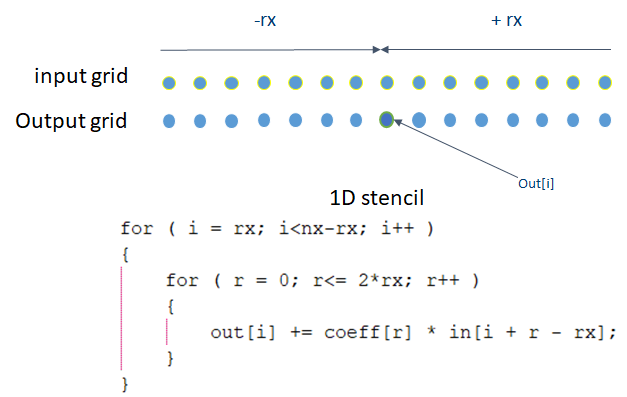}
\figcaption{A simple 1D $(2rx+1)$-point 1D stencil.}
\label{fig:stencil1D}
\end{figure}

\subsection{Stencil}
\label{sec:stencil}
Fig. \ref{fig:stencil1D} shows a simple 1D stencil pattern and the corresponding $C$ code snippet, where an output value is computed as the inner product of the coefficients and input data at $2*rx+1$ points, where $rx$ is the {\it radius} of the stencil. We call such a stencil a {\it $(2rx+1)$-point stencil}. In general, a stencil can be $1D$, $2D$, $3D$, $4D$ or even of higher dimension.

\begin{figure}[ht!]
\centering
\includegraphics[width=0.5\textwidth]{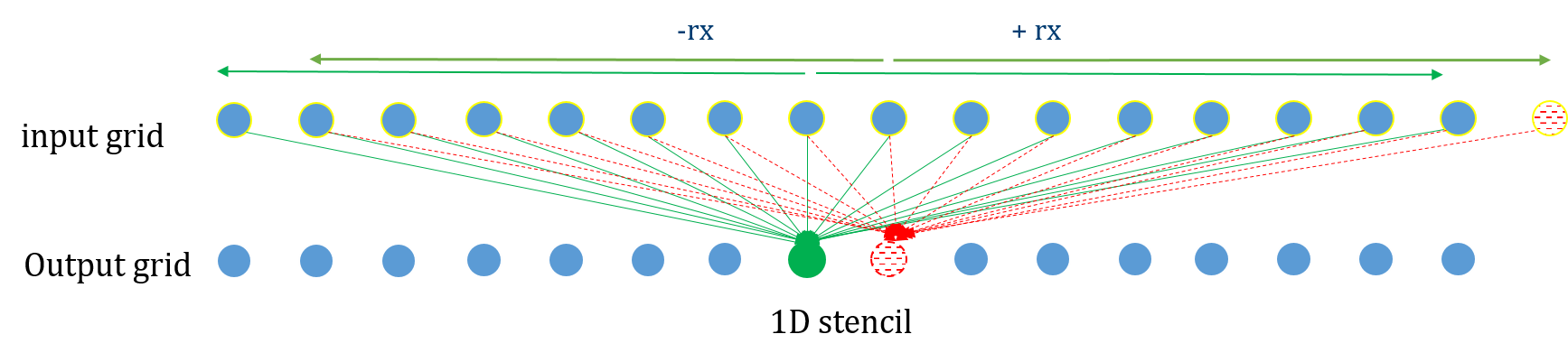}
\figcaption{
Data reuse opportunity in a 1D stencil. When computing the red cell in the output grid, all except one input data of the neighbouring green cell can be reused; Only one new input data needs to be loaded from memory.}
\label{fig:stencil1DDF}
\end{figure}

A stencil computation pattern allows loaded data to be reused both spatially and temporally. For the 1D stencil, as shown in Fig. ~\ref{fig:stencil1DDF}, each grid point can be loaded once, and then re-used for $2\times rx $ number of times without being re-loaded. The input data used by a worker computing a single output grid point can be reused by workers working on the neighboring grid points. For example, worker computing the red cell can reuse all input data loaded by worker computing the green cell except one in the leftmost one.

In general, any data point can be reused $2 * radius$ times along a dimension - provided we have enough on-chip storage to keep them around until they get reused. This allows the data to be loaded only once for the entire stencil computation.

Usually, the stencil computation as shown in Fig. ~\ref{fig:stencil1D} will be repeated many times. Every time step's output becomes the input of the next time step. To improve temporal data locality, the computations of multiple time steps can be combined into a pipeline, with I/O happening only at the beginning and end of the pipeline~\cite{Combined:Hamid, stencil:TYKTGC}.

\section{Mapping Stencils to a CGRA}
\label{mapping}
In this section, we present an algorithm to map a `star' stencil pattern to our target CGRA and discuss the challenges and design choices we made to overcome them. We start by discussing a 1D stencil mapping followed by a generalizable 2D mapping. We focus on a single time step iteration of the stencil. It is quite straightforward to extend our algorithm to multiple time-steps for a better temporal locality, and we leave that for future work.

\subsection{1D Stencil}
Without loss of generality, let us assume that the radius of the stencil \(rx\) is \(1\) for the $(2rx+1)$-point 1D stencil shown in Fig. \ref{fig:stencil1D}. Then, each output grid point \(out[i]\) would depend on \(3\) input grid points, $in[i-1]$, $in[i]$, $in[i+1]$. 

\begin{figure}[ht!]
\centering
\includegraphics[width=0.4\textwidth]{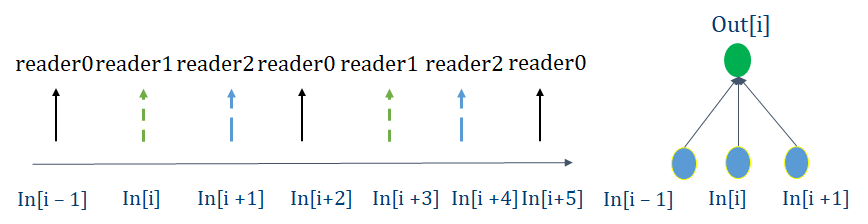}
\figcaption{For a simple 3-pt 1D stencil, the readers load data in an interleaved manner. Outputs from the same reader worker are shown in the same color/pattern.}
\label{fig:3Pt1Dstencil}
\end{figure}

We break the overall stencil computation into a pipeline of four stages: reading input, computing output, writing output, and synchronization. To exploit data and task parallelism, we employ $w$ ``workers'' to run each stage in parallel. ``Worker'' is a logical concept. A worker can be physically mapped to one or more PEs. 

Without loss of generality, lets assume that we have enough PEs to map \(3\) workers (\(w=3\)). However, the design is generalizable and works with any \(w\) including \(w=1\). Then we have a team of \(3\) workers to read data from memory in parallel, \(3\) workers to compute \(3\) consecutive output cells in parallel, \(3\) writers to write the output values to memory in parallel, and \(3\) synchronization workers working in parallel to ensure all writes are done in proper order. We explain the organizations of the workers below.

\underline{\bf Reader Workers. }
A reader consists of multiple PEs of the CGRA. We assume that the reader workers are loading data from the input grid in an interleaved manner as shown in Fig. \ref{fig:3Pt1Dstencil}, i.e., reader $0$ is reading $in[i-1]$, $in[i+2]$, $in[i+5]$, ..., reader $1$ is reading $in[i]$, $in[i+3]$, $in[i+6]$, ... and reader $2$ is reading $in[i+1]$, $in[i+4]$, $in[i+7]$, and so on. The input data read by the reader are fed to the compute workers which can forward the data to a neighbor compute worker to reuse the data instead of reloading them.

\underline{\bf Compute Workers. }
The compute workers also compute the output cells in an interleaved manner, i.e., worker $0$ computes $out[i]$, $out[i+3]$, $out[i+6]$, ..., worker $1$ computes $out[i+1]$, $out[i+4]$, $out[i+7]$, ..., and worker $2$ computes $out[i+2]$, $out[i+5]$, $out[i+8]$ and so on. To compute an output cell, each compute worker needs to compute $out[i] = coeff[0] * in[i-1] + coeff[1] * in[i] + coeff[2] * in[i+1]$, i.e., requires one MUL (multiplier) PE and two Fused MAC (multiply-and-add) PEs. 

\begin{figure}[htp]
\centering
\includegraphics[width=0.4\textwidth, height=1.5in]{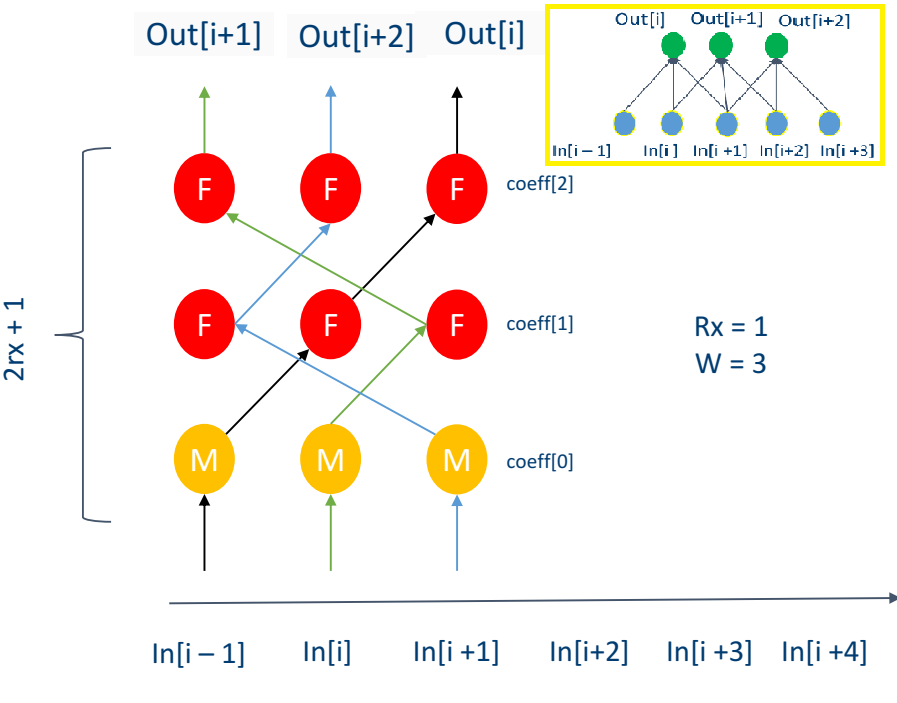}
\figcaption{PEs mapped on a physical rectangular CGRA grid.}
\label{fig:compute_block_all_worker}
\end{figure}

Fig. \ref{fig:compute_block_all_worker} shows the mapping of \(3\) compute workers on the CGRA spatial grid. Each worker has \(3\) PEs: one MUL (abbreviated as ``M" in the figure), and two Fused MACs (abbreviated as ``F"). In this mapping, the PEs in the same row share the same coefficient and the PEs in the same column receives data from the same reader. Each of the compute worker has additional data filtering PEs to filter out not-needed data from a reader worker. 

\begin{figure*}[h]
\centering
\includegraphics[width=0.24\textwidth, height=1.5in]{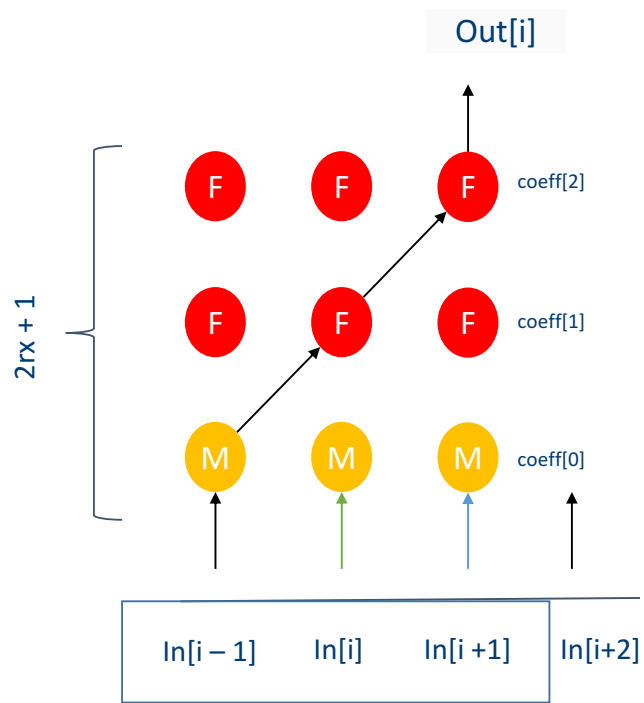}
\includegraphics[width=0.24\textwidth, height=1.5in]{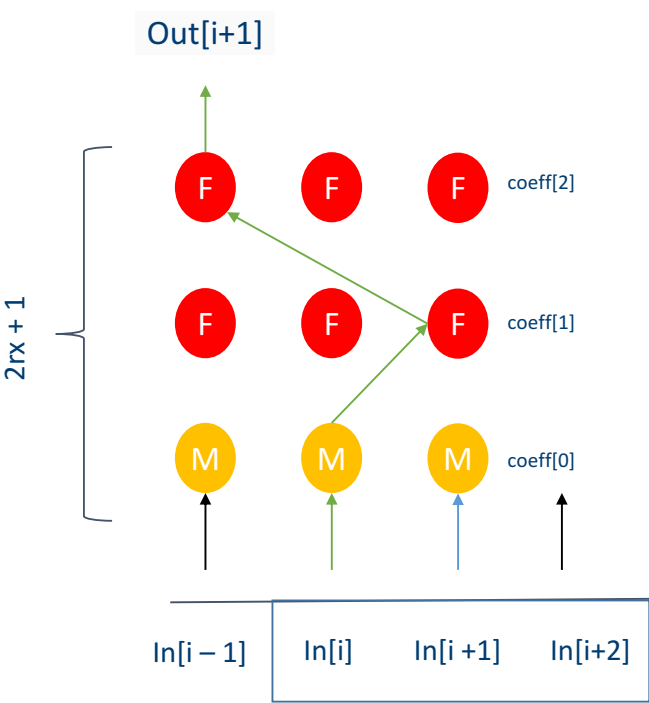}
\includegraphics[width=0.24\textwidth, height=1.5in]{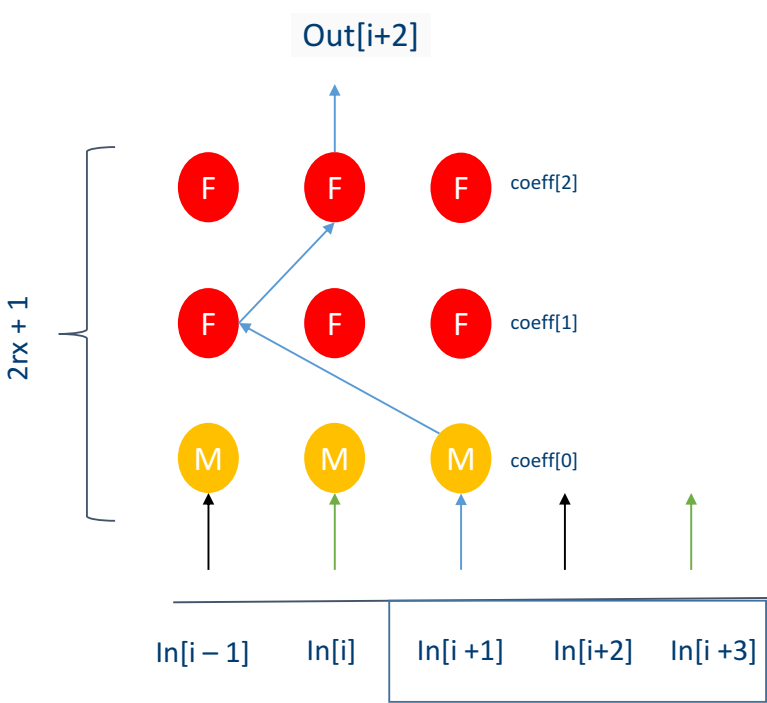}
\figcaption{Data-flow graph for the compute workers.}
\label{fig:compute_block}
\end{figure*}

To understand how it works, Fig. \ref{fig:compute_block} shows the dataflow for each compute worker side by side. The first sub-figure of Fig. \ref{fig:compute_block} shows how the MUL/MACs for compute worker $0$ are connected. Worker $0$ computes $out[i] = coeff[0] * in[i-1] + coeff[1] * in[i] + coeff[2] * in[i+1]$. Therefore, first, a MUL uses $in[i-1]$ (the output from reader worker $0$), multiplies it with a constant input $coeff[0]$, and sends the result to a MAC. The MAC uses $in[i]$ (the output from reader worker $1$), multiplies it with a constant input $coeff[1]$, and sends the result to a second MAC. The second MAC uses $in[i+1]$ (the output from reader worker $2$), multiplies it with a constant input $coeff[2]$, and gets the final result $out[i]$. The other two compute workers work similarly.

\underline{\bf Writer Worker. }
The writer worker $i$ takes the output of compute worker $i$ and stores that to memory. 

\underline{\bf Control Units. }There are control units attached to both reader and writer workers, generating addresses and row/column id corresponding to the load/store operations.

\underline{\bf Synchronization Worker.}
Each synchronization worker counts the number of stores from the corresponding writer worker it is tracking and triggers its own output when a certain number of stores are done. How many stores a store worker expects can be analytically counted, given the input grid size, the radius of the stencil, and the worker id. After all synchronization workers trigger their outputs, those outputs are combined to send a final ``done" signal to the host.

\begin{wrapfigure}{l}{0.2\textwidth}
\centering
\includegraphics[width=0.2\textwidth]{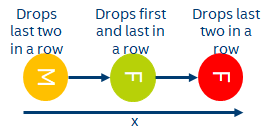}
\figcaption{Data drop Pattern.}
\label{fig:drop}
\end{wrapfigure}
\underline{\bf Data-filtering PEs. }
Each of the MULs and MACs has data filtering PEs to filter out `not-needed' data broadcasted by a reader worker in the same column. Imagine we have one compute- and reader- worker to compute the 3-point 1D stencil and the stencil grid size is $N$. The compute worker has one MUL and two MACs. In this scenario, the MUL will need to drop the last two input values of the input grid, the first MAC would need to drop the first and the last values of the row and the second MAC would drop the first two values as illustrated in Fig. \ref{fig:drop}. 

To filter data, one can use a bit generator to generate a bit pattern in the form of \(0^m1^n0^p\), e.g., \(00010\), and then use this bitstream to drop data when there is a $0$ and use the data when the bit is $1$. Here, $m$, $n$, and $p$ are constants that depend on the stencil pattern and the location of the MUL/MAC unit. In the above scenario, the MUL's filter will use a pattern $1^{(N-2)}00$, first MAC's filter will us a pattern $01^{(N-2)}0$ and last MAC's filter will use a pattern \(001^{(N-2)}\). These patterns can also be saved in scratchpad memory and shared among PEs that need the same filtering pattern.

Another way to do the filtering is to check the row id of the input data, and then drop it based on whether data with that row id is expected. For example, in the above mentioned case, the MUL's filter PEs will filter data if $row\_id > N-2$, the first MAC's filter PEs will filter out data if  $1 < row\_id \le N - 2$, the second MAC's filter PEs will filter out the data if $row\_id \le 1$.

\begin{figure*}[htp]
\centering
\includegraphics[width=0.8\textwidth]{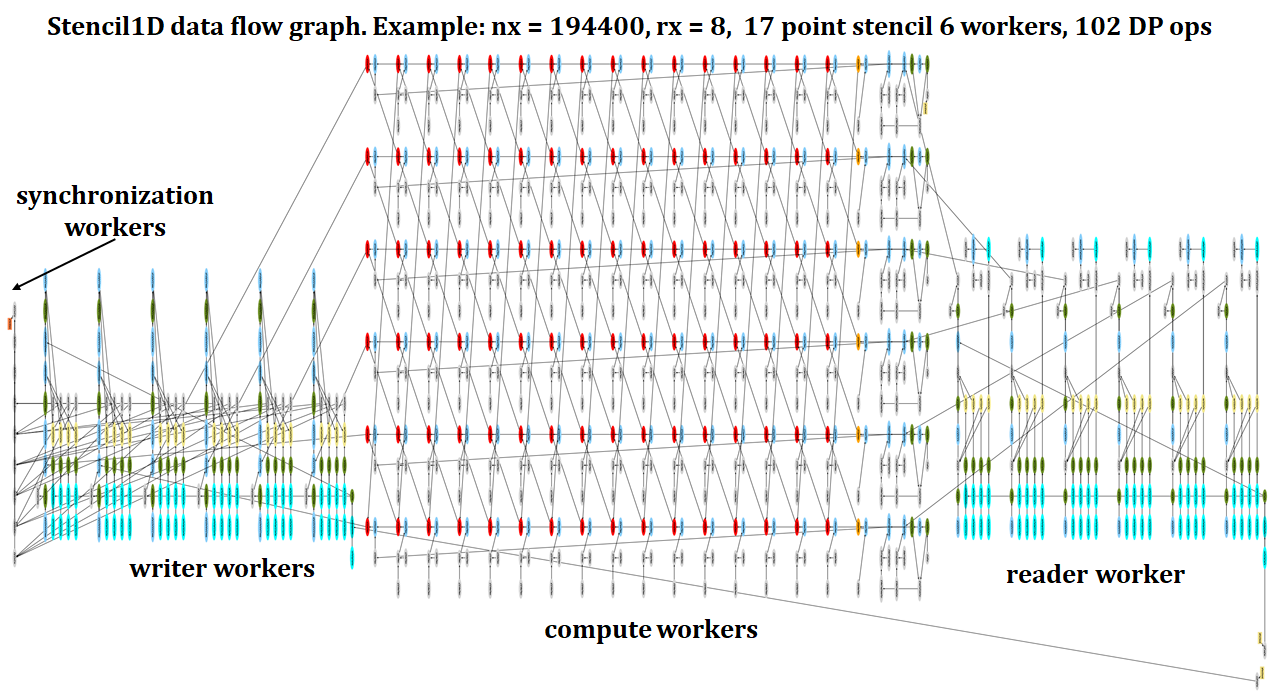}
\figcaption{Dataflow graph for a \(17\)-pt 1D stencil with \(6\) workers laid out on the CGRA grid.}
\label{fig:stencil1D_DF}
\end{figure*}

Fig. \ref{fig:stencil1D_DF} shows a complete Dataflow Graph for 1D stencil. In the graph, the light yellow ovals represent mux operation (op), the orange ovals represent MUL ops, the red ovals represents MAC ops, light blue ovals represent demux op, green ovals represent add op, cyan ovals represents addresses generators and indexes, gray oval represents any other PEs such as cmp, or, copy, and shift ops.

\subsection{2D Stencil}

\begin{wrapfigure}{l}{0.2\textwidth}
\centering
\includegraphics[width=0.2\textwidth]{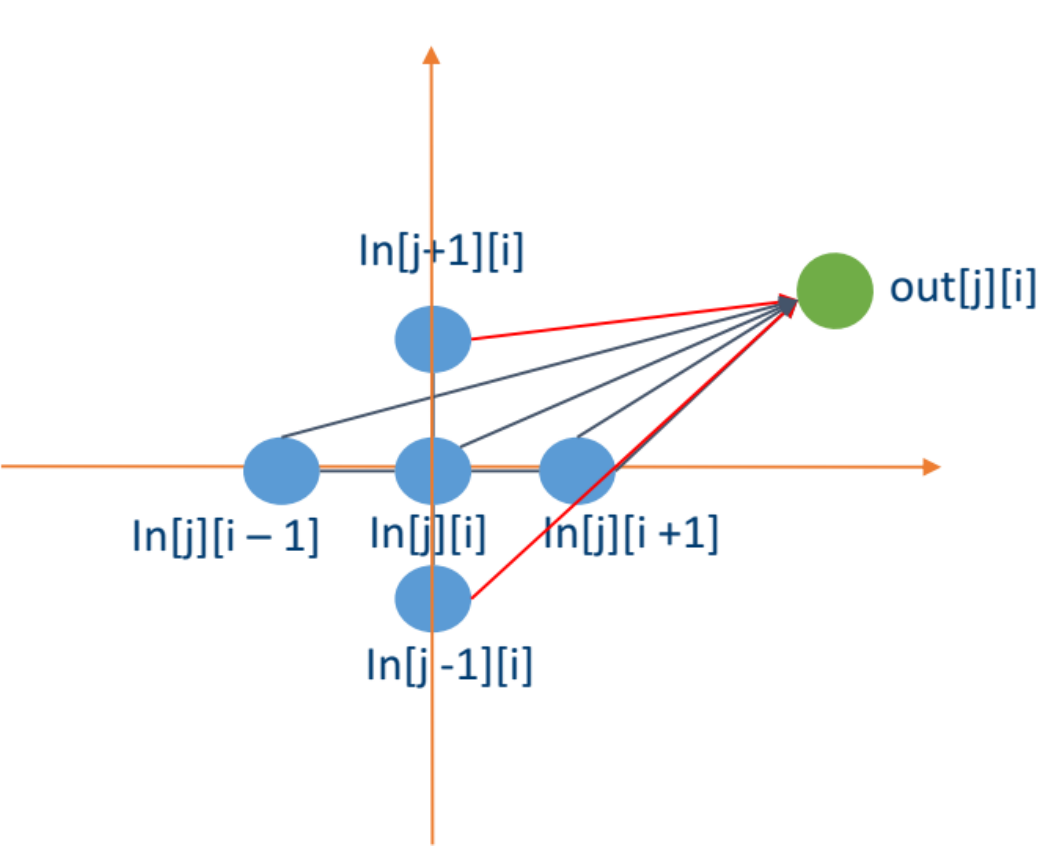}
\figcaption{A \(5\) point 2D Stencil.}
\label{fig:Stencil2D}
\vspace{-10pt}
\end{wrapfigure}
The input and output grids of a 2D steancil have two dimensions: $x$ and $y$. Each output grid point depends on a number of neighboring points from both dimensions. Figure \ref{fig:Stencil2D} shows a $5$-point 2D stencil where each \emph{out[j][i]} is computed using \emph{in[j][i]}, \emph{in[j][i-1]}, \emph{in[j][i+1]}, \emph{in[j-1][i]}, \emph{in[j+1][i]}.

We have explored multiple ways to implement a 2D stencil on the CGRA and in this section we describe one such mapping that is a natural extension of our stencil1D algorithm. This design can be extended to 3D stencil as well.

The stencil1D Algorithm can be naturally extended to 2D. In this case, the stencil contributions along the $x$ dimension \(coeff_x[0] * in[j][i-1] + coeff_x[1] * in[j][i] + coeff_x[2]\) can be computed following the same approach as stencil 1D. We also need to compute stencil contributions along the $y$ dimension \(coeff_y[0]*in[j-1][i] + coeff_y[2]*in[j+1][i]\) and add that to the partial sum along $x$ dimension to get the final value of \emph{out[j][i]}. The reader and writer workers remain the same. The control units controlling what to read and write need to be modified appropriately to read from or write to a 2D array.

\begin{figure}[htp!]
\centering
\includegraphics[width=0.5\textwidth]{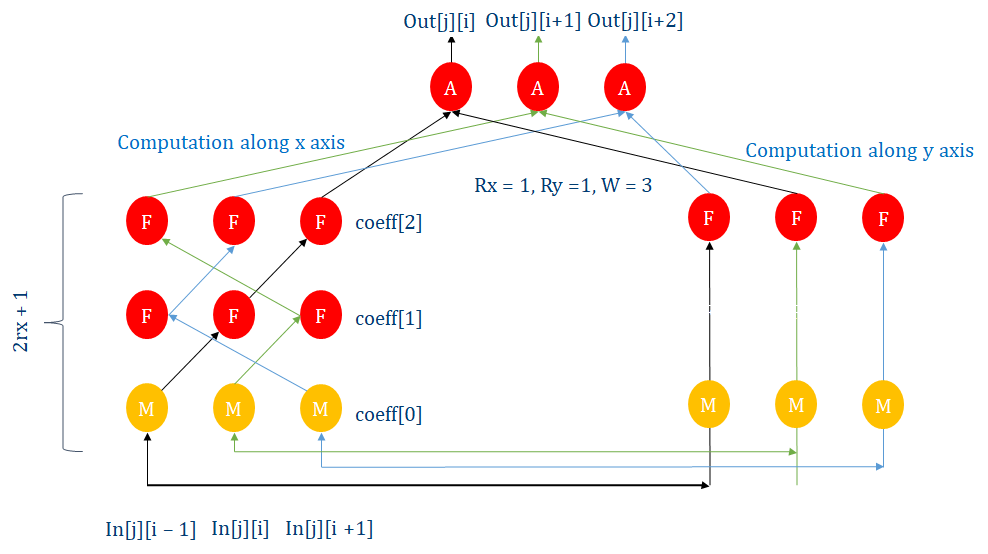}
\figcaption{A 5-point 2D Stencil.}
\label{fig:Stencil2D_compute_block}
\end{figure}

Figure \ref{fig:Stencil2D_compute_block} shows a simplified logical layout of \(w=3\) compute workers for the 2D $5$-point Jacobian stencil with radius along $x$ and $y$ dimensions, $rx=ry=1$. The dataflow diagram of the compute workers along $x$ dimension looks the same as stencil1D for obvious reason.

\underline{Compute workers along the $y$ dimension.} To compute the stencil contribution along the $y$ dimension, a chain of $2ry$ MUL/MAC is needed for each compute worker along the $y$ dimension. Unlike $x$ dimension, where each MAC/MUL in a compute worker's MAC chain accepts data from different readers, for $y$-dimension's compute workers, all MUL/MAC's input comes from only one particular reader worker's output. This is because the same reader worker reads all values of a given column. 

\underline{Reader workers for $y$ dimension.} We do not need separate reader workers to load values for $y$ dimension. The same reader workers that read values for $x$ dimension can feed the compute units (MAC/MUL chain) across $y$ dimension as well with proper data filtering and buffering. The goal is to read the data only once. It is important to note that, since the same reader workers are used to feed the compute workers in the $x$ and $y$ dimensions and since the computation along $x$ dimension does not start until the first $ry$ rows have been read and used by some compute workers for $y$ dimensions, we would need to filter out the data not needed by the $x$ compute workers. 

For the 2D $5$-point Jacobian stencil, compute worker $0$ in $y$ dimension should receive its data from reader worker $1$, since reader $1$ reads data $in[j-1][i]$, $in[j][i]$, $in[j+1][i]$ and all other values from column $i$. Compute worker $1$ in $y$ dimension should get its data from reader worker $2$ and compute worker $2$ in $y$ dimension should receive its input data from reader worker $0$. 

Finally, the partial sums produced by the $x$ and $y$ compute workers need to be added appropriately to produce the final output value as shown in Figure \ref{fig:Stencil2D_compute_block}.

\underline{\bf Mandatory Buffering.} The first $ry$ rows are used only to compute the $y$ contributions of this stencil. The last MAC PE that computes the $y$ contribution from the in[j+ry-1][i]th input point would need to wait for $2ry$ rows to be read before it can start computing. Therefore, a sufficient amount of buffering/storage is needed in the input and output queues of those PEs are needed to tolerate this wait time and avoid deadlock situation. The goal is to keep $2ry*x\_dim$ data all the time inside the queues to avoid reading data more than once. That way, when we walk along the $y$-dimension to compute any row of the output, we can reuse all except the bottom-most of the last $2ry$ rows of data read from input. We would still need to read one new row of input grid for each new row of the output. 

\begin{wrapfigure}{l}{0.2\textwidth}
\centering
\includegraphics[width=0.2\textwidth]{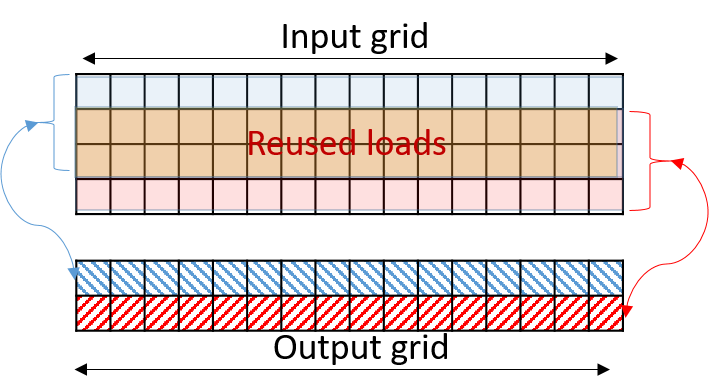}
\figcaption{Data reuse opportunities in a 5-point 2D Stencil.}
\label{fig:Datareuse_Stencil2D}
\end{wrapfigure}

\underline{Blocking.} The length of the $2ry$ rows that can be kept inside the CGRA queues/PE storage is limited by the amount of on-fabric storage. Therefore, if $x\_dim$ is very large, we would need to block the stencil grid into vertical strips so that $2ry*block\_size$ size data can fit in the available storage of the given CGRA. This is a variation of \emph{strip mining} \cite{tithi2014exploiting}. Figure \ref{fig:Datareuse_Stencil2D} hows how data can be reused in 5-point 2D Stencil. The red output cells are computed using pink input cells and the blue output cells are computed using the brown and blue input cells and the brown-colored cells are reused between two rows.

\begin{figure*}[ht]
\centering
\includegraphics[width=0.7\textwidth]{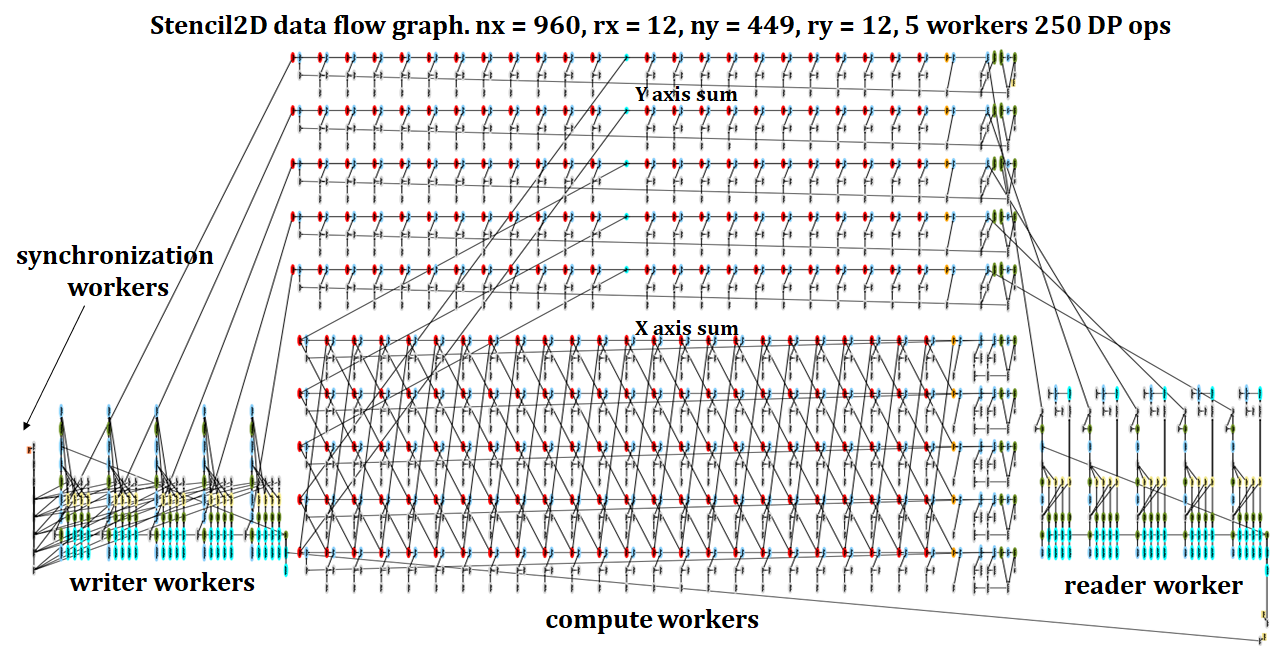}
\figcaption{Complete Dataflow Graph for 2D stencil.}
\label{fig:stencil2D_full_graph.pdf}
\end{figure*}

A complete stencil 2D graph for a five worker, \(49\)-pt 2D stencil with $rx=ry=12$ is shown in Figure \ref{fig:stencil2D_full_graph.pdf}.

\section{Algorithm with Temporal Locality}
We mentioned at the beginning of section \ref{mapping} that in the current work, we compute only single time-step of the stencil on the CGRA during a single kernel call since for our usecase other kernels need to be applied over the stencil grid before calling the stencil kernel again. However, there are stencil applications where multiple time-steps can be computed as once without applying other kernels on the stencil grid. In those cases, it is possible to exploit temporal locality by loading data for time-step $t$ and computing the next $k$ time-steps without storing intermediate data to the main memory. Such optimization of reusing data from cache or internal memory provides performance gains both for GPU \cite{Stencil2014GPU} and for FPGAs \cite{Combined:Hamid}.

To enable computation of multiple time-steps on our target CGRA, we would have to deploy workers along the time dimension as well.

Suppose we have our 2D stencil and we want to compute two time-steps in a pipelined fashion. To extend the original stencil2D algorithm to compute two time-steps in parallel, we would need to add another layer of compute workers for time step $t+1$. These compute workers would not need separate reader-workers: they would receive their input from compute workers computing time-step $t$ directly by connecting output of one PE to the input of another PE. The writer workers that used to get inputs from the output of the compute workers at time step $t$ would now instead receive inputs from the output of computer workers working on time-step $t+1$. We leave a full-fledged implementation of this algorithm for the future.

Researchers have used recursive divide-and-conquer wavefront algorithms on multicores and manycores to implement high-performance stencils that enable efficient data reuse on cache and exploit both temporal and spatial locality \cite{stencil:TYKTGC}. To exploit temporal locality, a recursive divide-and-conquer algorithm can be used to generate small stencil subtasks which can then be offloaded to a CGRA. If multiple CGRA chips are available, a hybrid CPU + CGRA algorithm can be designed where multiple CPU cores sharing the same last level cache can offload independent stencil tasks to the CGRAs. This approach would allow creating small stencil subtasks that fit the CGRA fabric and execute them in a way that is cache efficient from the perspective of the big cores and can also fully reuse the on-fabric data while running on CGRA.

\section{Creating Data-flow Graphs}
An algorithm for a CGRA is represented as a data-flow graph (DFG) where nodes represent instructions and edges represent producer-consumer relationships between the instructions. The instructions used to program the CGRA creates the DFG. Creating data-flow graph (DFG) for a complex stencil could be challenging. To create the DFGs in a user-friendly and scalable way, we created a High-Level Domain Specific Language (DSL) tool that provides essential APIs to add PEs and connect their inputs and outputs to create each building block (pipeline stage: control units-, reader-, compute-, writer- and synchronization- workers) parametrically. The tool automatically connects the operations internally based on the input/output names of each operation and creates the DFG accordingly. The tool then emits a high-level assembly program for the created DFG which can also be visualized using the Graphviz dot tool and can be saved in files. Using this tool, we essentially coded a general multi-dimensional stencil generator to generate the high-level assembly programs of stencils with different radius, worker and dimensions. 
\section{Roofline Performance Analysis}
In this section, we show the roofline performance analysis of our stencil implementations on the target CGRA. The roofline analysis helps us to choose the optimal number of workers for a given stencil based on its arithmetic\_intensity (flops per byte ratio) and the compute and bandwidth capacity of the target CGRA. 

We assumed the following about the target CGRA: clock speed = 1.2 GHz, Number of MACs=$256$, and Bandwidth (BW) = $100$GB/s. Therefore, the maximum GFLOPS that can be obtained from this CGRA is $2*256*1.2$G=$614$ GFLOPS if not limited by bandwidth. 

If we assume that the memory bandwidth is $X$ GB/s, then the expected executed flops with the given arithmetic\_intensity (flops per byte ratio) would be X $*$ arithmetic\_intensity GFLOPS. If we have $Y$ MAC PEs, we could fit $Y\over \#MACs\_per\_workers$ workers for a given stencil. The minimum of these two numbers dictates what could be the best performance achievable on a CGRA.

\underline{\bf 1D Stencil}
For stencil 1D, we used a radius $8$ (i.e. rx = $8$) $17$-pt 1D stencil on a grid size of $194400$. The arithmetic\_intensity for such a stencil is $(16*2+1)*(194400-16)/((194400+194400)*8)=2.06$. The expected Gflops with this arithmetic\_intensity is $100*2.06 = 206$. If we use $6$ workers with $(6* (16 MAC + 1 MUL)) = 102$ Double precision (DP) flops at a clock frequency of $1.2$ GHz, that would demand $6*16*2*1.2 + 6*1.2=237$ GFLOPS. Therefore, the roofline model suggests that use of $6$ workers should be good enough to saturate the achievable memory bandwidth.

\begin{figure}[ht!]
\begin{tabular}{cc}
\centering
\includegraphics[width=0.24\textwidth]{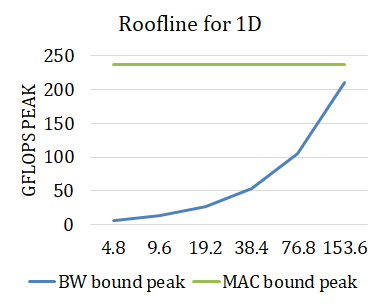}&
\includegraphics[width=0.24\textwidth]{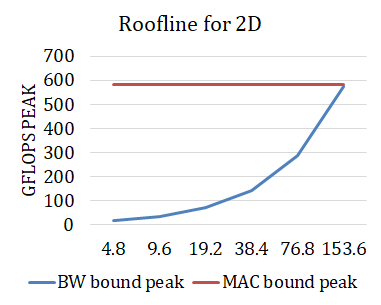}
\end{tabular}
\figcaption{Roofline model for stencil1D and stencil2D.}
\label{fig:roofline_stencil2D}
\end{figure}

\underline{\bf 2D Stencil}
We used a 2D stencil structure from the oil/gas seismic simulation with a grid size of $960\times449$ and $rx=ry=12$. The arithmetic\_intensity for this stencil is $(48*2+1)*((449 - 24)*(960-24))/((2*(960*449))*8) ~=~ 5.59$. Therefore, the peak GFLOPS limited by the memory bandwidth is $100*5.59 = 559$. Since each worker requires $49$ double precision ops ($48$ MAC and $1$ MUL), we can only fit $5$ workers in one CGRA. Therefore, the GFLOPS limited by number of double precision PEs running at $1.2$ GHZ frequency is $1.2 * (48 * 2* 5 + 5) = 582$. Based on the roofline (Fig. \ref{fig:roofline_stencil2D}) model, the peak FLOPS achievable here is $559$ GFLOPS.

\section{GPU Implementation and Optimizations}
In this section, we describe the GPU implementation and optimizations we used to get the best performance out of the GPU. 

The stencil GPU code that we used for performance comparison has been adopted from \cite{Micikevicius} and subsequently optimized based on \cite{Stencil2014GPU} (except the temporal blocking part) to stick with the single time-step use-case.  We then applied additional optimizations as described below.

\underline{\bf{Usage of constant cache:}} We used the GPU constant cache to store the stencil coefficients. This means that coefficients are broadcasted from the constant cache to the threads in the warp, reducing pressure on the GPU Shared Memory (SMEM).

\underline{\bf {Explicit use of Shared memory:}} The original GPU implementation where each CUDA thread processes one output cell relies on automated L1 caching to achieve higher performance. On V100, L1 and SMEM are combined into a 128KB block and the explicit usage of SMEM in the CUDA kernel provides around $10$\% more performance.

The main bottleneck in the stencil GPU implementation when using shared memory is the high number of redundant loads. Given that each thread computes one output cell and that threads don’t communicate, there is redundant traffic from SMEM, because each thread reads neighboring cells. On top of that, bank conflicts are common for reading neighbors on one dimension. The SMEM read latency is more than \(25\) clocks so at least \(25\) warps are required to fully hide this latency. 

The number of resident warps is limited by the SMEM size, which is, 96KB. This space is used less efficiently as the stencil radius increases because we need to keep the $2*radius$ elements in the shared memory. This leads to less than the optimal number of eligible warps per cycle. Additionally, copying data from global memory to SMEM requires synchronization, which causes an additional penalty. Due to the above-mentioned inefficiencies, we can’t use the peak SMEM bandwidth and we observed around $60$\% utilization during the runs. The overall GFLOPs for this implementation was $1900$GFLOPS for the 2D stencil.

\underline{\bf {Register Caching and Blocking:}} To alleviate some of the pressure on the SMEM, we used ``register caching" where each warp computes essentially a 2D block of the stencil. In the best implementation, we split the stencil computation into blocks that are $32 \times 8$ elements (grid points) and each block was assigned to 1 warp. Each thread in the warp computed $8$ elements, instead of 1 as in the SMEM implementation. In the first step of the optimized kernel, the 2D blocks are copied to SMEM. Then, each thread reads the neighboring elements and the target element into registers and computes one output value. For the next output value, only one SMEM read on each dimension is required, along with a circular shift in registers. Given that each thread computes 8 outputs, the number of redundant SMEM reads is reduced by a factor of 8. 

The implementation has a vertical and a horizontal pass. Depending on the storage format in SMEM (column or row-major), one of the passes generates bank conflicts as in the exclusive SMEM implementation, but the impact will be lower given the reduced count of redundant reads. In this implementation, the bottleneck is the register file size, which limits the number of resident warps. However, given that the SMEM accesses count is lower, and that FP64 instructions are generally 8 cycles which can be hidden with 8 warps, this bottleneck still allows for better efficiency than the previous implementation. For the register-reuse CUDA kernel, we obtained $2300$ GFLOPS.

This optimized GPU implementation obtained $77\%$ of the roofline estimated peak performance on V100 for the same stencil as used in \cite{Stencil2014GPU} with a grid size of $384\times384\times128$ for single precision and $80\%$ of the peak for double-precision data. As mentioned before, the GPU shared memory blocking has been used to enhance shared memory data reuse and to ensure that data is fetched from global memory as few times as possible. The GPU constant memory has been used to store read-only data for uniform access across threads in the warp. With the increase of arithmetic intensity \emph{flops per byte ratio}, the efficiency of the stencil dropped on V100. For example: for a 3D stencil with $rx=ry=rz=8$ and grid size $384\times384\times384$, for single-precision data, the efficiency (\% of peak) dropped to $56\%$ and for $rx = ry = rz = 12$ and grid size $512\times512\times512$, single precision data, the efficiency dropped to $36\%$.

\section{Experimental Result}
In this section, we discuss the performance results of our stencil implementations on a modified version of a previously proposed CGRA\cite{TIA:micros}. We used a cycle-accurate simulator to simulate CGRA PEs, scratchpads, private cache, shared cache, and communication network across PEs and to scratchpad, cache, and memory. All experiments have been done on one CGRA which then got extrapolated to estimate performance on $16$ CGRA tiles. 

A cycle-accurate simulation shows our stencil 1D implementation can reach $91$\% of the peak performance for Grid = $194400$, rx = $8$, and $77$\% of the peak performance for stencil 2D for Grid = $960$ x $449$, rx = $12$, ry = $12$ on the CGRA (see Table \ref{fig:stencil_performance}). We observed more conflict misses in the cache for stencil 2D implementation than stencil 1D. 

\subsection{GPU vs CGRA}
We compare the performance of our stencil implementations on $16$ CGRA units with that of an Nvidia V100 GPU, because, $16$ CGRA units should occupy the same chip area as that of a V100 GPU.

Table \ref{fig:stencil_performance} shows performance comparison of $16$ CGRA tiles with Nvidia V100 processor. To estimate peak performance on V100, we assumed the peak copy bandwidth on V100 is $850$ GB/s. For CGRA, the peak bandwidth has been assumed to be $100 * 16 = 1600GB/s$. For Stencil 2d, with $5.59$ flops/byte and with a peak bandwidth of $850$ GB/s, peak roofline performance is $4.8$ TFLOPS. For the target problem size, we achieved $2.3TFLOPS$ on V100. With a larger stencil grid size, performance increased from $2.3$TFLOPS to $2.85$TFLOPs.

The result in Table \ref{fig:stencil_performance} shows that $16$ CGRA tiles are $1.9\times$ faster than V100 for stencil1D and $3\times$ faster for stencil2D. We observed that the V100 performance gets closer to the roofline peak as we decrease the arithmetic\_intensity of the stencil by reducing the radius. For example, a 2D stencil with $rx=ry=2$ achieved $87$\% of the estimated peak for the same grid size.

\begin{table}[htp]
\centering
\resizebox{0.5\textwidth}{!}{
    \begin{tabular}{|c|c|c|}
    \toprule
          & \multicolumn{2}{c|}{\textbf{Speedup wrt CGRA}} \\
    \midrule
    \textbf{Stencil 1D (Grid = 194400, rx = 8)} & \textbf{CGRA} & \textbf{V100} \\
    \midrule
    \textbf{Normalized GFLOPS } & 1.9   & 1 \\
    \midrule
    \textbf{\% Peak} & 91\%  &  90\% \\
    \midrule
    \textbf{Stencil2D (Grid = 960x449, rx = 12, ry = 12)} & \textbf{CGRA} & V100 \\
    \midrule
    \textbf{Normalized GFLOPS } & 3.03  & 1 \\
    \midrule
    \textbf{\% Peak} & 78\%  & 48\% \\
    \bottomrule
    \end{tabular}%
    }
\tabcaption{Comparative analysis of stencils on CGRA and GPU.}
\label{fig:stencil_performance}
\end{table}

Fundamentally, the CGRA has an advantage over GPGPUs - the much higher speed internal fabric. We estimated that the PE to PE communication on CGRA is approximately $6\times$ faster compared to V100 register to shared memory communication time, and $3\times$ faster than register to register time on V100. The size of register file or shared memory limits the FLOP efficiency on V100 by limiting the number of eligible threads per cycle. 

\section{Related Work}
As FPGAs have been gaining traction as accelerator alternatives, researchers have implemented different types of stencils on FPGAs \cite{Luzhou,Yamamoto,Jia,Combined:Hamid}. FPGAs also have a large number of internal registers and thus, an optimized FPGA stencil implementation can also benefit from internal data reuse as we get on CGRA. Therefore, FPGAs can have similar performance and energy improvement potentials for stencils as on CGRA. Unfortunately, the traditional way of programming, simulating, debugging and testing on FPGA is quite difficult and time-consuming and thus, have rarely used in actual scientific computation environments irrespective of their superior performance potential compared to multicore CPUs and GPUs. 

For CGRA, we used a C based tool to create the data-flow graphs. One could potentially use an alternative python-based tool as well to do the same. Since we work at instruction and meta-instruction level granularity; it is relatively easy to design, program, simulate, test and debug compared to FPGAs. 

Recently, researchers have used high-level programming language, OpenCL to overcome those disadvantages for FPGA \cite{Combined:Hamid,Waidyasooriya,Jia,Wellery}. However, the reported performance in those research indicates lower efficiency compared to what we see on our target CGRA. 
\section{Conclusion}
In this paper, we show a unique and generalizable way of mapping one- and multi-dimensional stencil computations to a CGRA, fully exploiting the data reuse opportunities and parallelism. Our simulation experiments demonstrate that these mappings are efficient and enable the CGRA to outperform state-of-the-art GPUs with a highly optimized cuda implementation. We demonstrate the ability of a CGRA to accelerate stencils. We show that it is possible to get more than $3\times$ performance improvement in runtime compared to a highly optimized CUDA based implementation run on a high-end Nvidia GPU processor (V100).
This acceleration essentially comes from the exploitation of fine-grained parallelism and the dramatic reduction of memory reads and writes due to efficient data reuse on the chip. Our experiments show that the CGRA can be a better alternative for accelerating stencil computations compared to GPUs. 
\bibliographystyle{IEEEtran}
\bibliography{IEEEabrv,Stencil_H2RC}
\end{document}